\documentclass{article}
\newcommand{\bq}{\begin{equation}}
\newcommand{\eq}{\end{equation}}
\newcommand{\ba}{\begin{eqnarray}}
\newcommand{\ea}{\end{eqnarray}}

\newcommand{\nl }{ \nonumber  }

\newcommand{\p}{\partial}

\newcommand{\h}{\hspace{.5cm}}
\newcommand{\s}{\sigma}

\newcommand{\la}{\lambda}

\title{On the Open String Ending on D-brane}
\author{ P. Bozhilov}                       

\begin{document}                            

\maketitle                                  

We obtain background independent solutions for an open string
ending on D-brane, in variable external fields. Explicit solution
of the boundary conditions is given for background metric and
NS-NS two-form gauge field, depending on the coordinates of the
transverse to the D$p$-brane directions. Extension of the
constraint algebra is proposed and discussed from both Hamiltonian
and Lagrangian approach viewpoint.

\section{Introduction}
Obtaining exact solutions of the nonlinear probe string equations
of motion and constraints in variable external fields is by all
means an interesting task with many possible applications. One
such application is connected with the recent investigations of
the open string - D-brane system in non-constant background fields
\cite{CHK99}-\cite{ZH03}, in which noncommutative Yang-Mills and
noncommutative open string theories can arise on the D-brane
worldvolume. Of course, in this case, one is forced to use
different approximations in order to get explicit results. That is
why, it is interesting to see to what extent our knowledge about
the existing {\it exact} string solutions can help us in
considering this dynamical system.

In this paper, we obtain background independent solutions of the
open string equations of motion and constraints in non-constant
background metric and NS-NS two-form gauge field. Explicit
solution of the boundary conditions for the open string ending on
D-brane is given for background metric and NS-NS fields, depending
on the coordinates of the transverse to the D$p$-brane directions.
Then we check on an example their compatibility with the solutions
of the equations of motion and constraints. After that, we
reinterpret the conditions for existence of such solutions as a
set of constraints and compute their Poisson bracket algebra. The
consequences from the Lagrangian approach viewpoint are also
given.

\section{The Open String - D-brane System in String Theory Background}
The action for an open string ending on a D$p$-brane, in the
presence of background gravitational and NS-NS 2-form field, can
be written as \ba\label{asd} &&S_1=-\frac{T}{2}\int d^{2}\xi
\left[ \sqrt{-\gamma}\gamma^{mn} \p_m X^M\p_n X^N g_{MN}(X)\right.
\\ \nl &&- \left.\varepsilon^{mn}\p_m X^M\p_n X^N B_{MN}(X)\right]
-\frac{T}{2}\int d^{2}\xi\varepsilon^{mn}\p_m Y^\mu\p_n Y^\nu
F_{\mu\nu}(Y),\\ \nl &&\p_m =\p/\p\xi^m,\h
\xi^m=(\xi^0,\xi^1)=(\tau,\s),\\ \nl &&m,n=0,1,\h M,N=0,1,...,D-1,
\\ \nl &&F_{\mu\nu}=\p_\mu A_\nu - \p_\nu A_\mu,\h
\mu,\nu=0,1,...,p,\ea where $T=\left(2\pi\alpha'\right)^{-1}$ is
the (fundamental) string tension, \ba\nl G_{mn}(X) = \p_m X^M\p_n
X^N g_{MN}(X),\h B_{mn}(X)=\p_m X^M\p_n X^N B_{MN}(X)\ea are the
pullbacks of the background metric and antisymmetric NS-NS tensor
to the string worldsheet, $\gamma$ is the determinant of the
auxiliary metric $\gamma_{mn}$, $Y^\mu(\xi)$ are the coordinates
on the D-brane, and $A_\mu$ is the $U(1)$ gauge field living on
the D-brane worldvolume. In {\it static gauge} for the D-brane,
one identifies $Y^\mu$ with the string embedding coordinates
$X^\mu(\xi)$. Then the action (\ref{asd}) acquires the form
\ba\label{asB} S_2=&-&\frac{T}{2}\int d^{2}\xi \left[
\sqrt{-\gamma}\gamma^{mn} \p_m X^M\p_n X^N g_{MN}(X)\right.\\ \nl
&-& \left.\varepsilon^{mn}\p_m X^M\p_n X^N B'_{MN}(X)\right], \ea
where \ba\nl B'_{MN}= B_{MN} - \delta_M^\mu \delta_N^\nu
F_{\mu\nu}.\ea

Varying (\ref{asB}) with respect to $X^M$ and $\gamma_{mn}$, one
obtains the equations of motion \ba\label{em}
&&-g_{LK}\left[\p_m\left( \sqrt{-\gamma}\gamma^{mn} \p_n
X^K\right) + \sqrt{-\gamma}\gamma^{mn}\Gamma^{K}_{MN}\p_m X^M\p_n
X^N \right]
\\ \nl &&= \frac{1}{2}H_{LMN}\varepsilon^{mn}\p_m X^M\p_n X^N, \ea and
the constraints \ba\label{con1}
\left(\gamma^{kl}\gamma^{mn}-2\gamma^{km}\gamma^{ln}\right)\p_m
X^M\p_n X^N g_{MN}\left(X\right)=0,\ea where \ba\nl\Gamma^{K}_{MN}
=g^{KL}\Gamma_{LMN} = \frac{1}{2}g^{KL}\left(\p_Mg_{NL} +
\p_Ng_{ML} - \p_Lg_{MN}\right)\ea is the symmetric connection
compatible with the metric $g_{MN}$ and \ba\nl H_{LMN}=\p_L
B_{MN}+\p_M B_{NL}+\p_N B_{LM}\ea is the $B_{MN}$ field strength.
The field $F_{\mu\nu}$ does not enter the equations of motion,
because $dF\equiv 0$.

The field $B'_{MN}$ explicitly appears in the expressions for the
generalized momenta \ba\label{gm} P_M =
-T\left(\sqrt{-\gamma}g_{MN}\gamma^{0n}\p_n X^N - B'_{MN}\p_1 X^N
\right),\ea and in the boundary conditions \ba\label{mbcs}
&&\left[\sqrt{-\gamma}g_{M\nu}\gamma^{1n}\p_n X^\nu +B'_{M\nu}\p_0
X^\nu\right]_{\s=0,\pi}=0,\\ \label{dbcs}
&&X^a(\tau,0)=X^a(\tau,\pi)=q^a,\h a=p+1,...,D-1.\ea Here we have
split the coordinates $X^M$ into $X^\mu$ and $X^a$, and have
denoted the location of the D-brane with $q^a$.

From now on, we will work in the gauge $\gamma^{mn}=constants$. We
note that $\gamma^{mn}=\eta^{mn}=diag(-1,1)$ correspond to the
commonly used {\it conformal gauge}.

\subsection{Solutions of the Equations of Motion}
First of all, we will try to find {\it background independent}
solutions of the equations of motion of the type \ba\nl
X^M(\xi)=F^M(a_n\xi^n),\h a_n = constants.\ea  It turns out that
such solutions exist when $\gamma^{mn}a_m a_n =0$. This leads to
\ba\label{uu} X^M(\xi)=F^M_\pm(u_\pm),\h
u_\pm=-\frac{1}{\gamma^{00}}
\left(\gamma^{01}\pm\frac{1}{\sqrt{-\gamma}}\right)\xi^0+\xi^1,\ea
or \ba\label{vv} X^M(\xi)=F^M_\pm(v_\pm),\h
v_\pm=\xi^0+\frac{1}{\gamma^{11}}
\left(-\gamma^{01}\pm\frac{1}{\sqrt{-\gamma}}\right)\xi^1,\ea
where $F^M_\pm$ are arbitrary functions of their arguments. The
main consequence of the obtained result is that for {\it arbitrary
background fields} there exist only one solution, $F^M_+$ or
$F^M_-$, but not both at the same time, in contrast with the flat
space-time case. In other words, we have only chiral background
independent solutions of the string equations of motion. On the
other hand, it must be noted that these are not solutions of the
constraints (\ref{con1}). Taking a linear combination of the two
independent constraints, we can arrange one of them to be
satisfied, but the other one will give restrictions on the metric.
However, it can be shown that in the zero tension limit, the
background independent solutions of the string equations of motion
are also solutions of the corresponding constraints. Moreover,
this result extends to arbitrary {\it tensionless} $p$-branes
\cite{B_PRD60}. It corresponds to the limit $(-\gamma)^{-1/2}\to
0$, taken in the expressions for $u_\pm$ and $v_\pm$. Let us also
note that in conformal gauge, the obtained string solutions
$F^M_\pm(u_\pm)$ and $F^M_\pm(v_\pm)$ reduce to the solutions
$X^M_\pm(\s\pm\tau)$ and $X^M_\pm(\tau\pm\s)$ for left- or
right-movers.

Our next step is to search for {\it non-chiral} solutions of the
string equations of motion and constraints, i.e. solutions of the
type \ba\label{ncs} X^M(\xi^m)=F^M_+(w_+) + F^M_-(w_-),\ea where
$w_\pm=u_\pm$ or $w_\pm=v_\pm$. Putting (\ref{ncs}) in the
equations of motion (\ref{em}), we obtain the conditions for the
existence of such solutions: \ba\label{ncsc}
\left(2\Gamma_{L,MN}+H_{LMN}\right)\frac{dF^M_+}{dw_+}
\frac{dF^N_-}{dw_-}=0.\ea For simplicity, we will consider the
case when $g_{MN}$ and $B_{MN}$ depend on only one coordinate, say
$r$ \footnote{In the next subsection, the coordinate $r$ will be
associated with the radial coordinate in the directions transverse
to the D$p$-brane, on which the open string ends.}, and will give
the results in conformal gauge.

In our first example, we fix all string coordinates $X^M$ except
$X^0$ and $r$ (the remaining coordinates are denoted as
$X^\alpha$). Then the conditions (\ref{ncsc}) and constraints
(\ref{con1}) reduce to the system of equations \ba\nl &&\p_r
g_{00}[(\p_0 X^0)^2 - (\p_1 X^0)^2] - \p_r g_{rr} [(\p_0 r)^2 -
(\p_1 r)^2]=0, \\ \nl &&\p_r B_{0\alpha}(\p_0 X^0\p_1 r-\p_1
X^0\p_0 r)+\p_r g_{0\alpha} (\p_0 X^0 \p_0 r-\p_1 X^0 \p_1 r)\\
\nl &&+\p_r g_{r\alpha}[(\p_0 r)^2-(\p_1 r)^2]=0,\\ \nl &&\p_r
g_{00}(\p_0 X^0\p_0 r-\p_1 X^0\p_1 r)+ \p_r g_{0r}[(\p_0 r)^2 -
(\p_1 r)^2]=0, \\ \nl &&g_{00}[(\p_0 X^0)^2 + (\p_1 X^0)^2] +
g_{rr}[(\p_0 r)^2\\ \nl &&+ (\p_1 r)^2]+ 2g_{0r}(\p_0 X^0\p_0
r+\p_1 X^0\p_1 r)=0,\\ \nl &&g_{00}\p_0 X^0 \p_1 X^0+g_{rr}\p_0 r
\p_1 r+ g_{0r}(\p_0 X^0\p_1 r+\p_1 X^0\p_0 r)=0.\ea Among the
nontrivial solutions of the above system, there exist the
following non-chiral ones: \ba\label{s01}
&&\frac{g_{rr}}{g_{00}}=\frac{\p_r g_{rr}}{\p_r g_{00}}=
-\left(\frac{\p_r g_{r\alpha}}{\p_r B_{0\alpha}}\right)^2,\h
g_{0r}=0,\\ \nl &&\p_r B_{0\alpha}\p_0X^0=\p_r g_{r\alpha}\p_1
r,\h \p_0 X^0 \p_0 r=\p_1 X^0 \p_1 r;\ea \vspace{.5cm} \ba\nl
&&\frac{\p_r g_{r\alpha}}{\p_r g_{0\alpha}}= \frac{\p_r
g_{0r}}{\p_r g_{00}},\h \p_r g_{rr}=\frac{(\p_r g_{0r})^2}{\p_r
g_{00}},\h g_{rr}=\left(2g_{0r}-g_{00}\frac{\p_r g_{0r}}{\p_r
g_{00}}\right) \frac{\p_r g_{0r}}{\p_r g_{00}},\\ \label{s02}
&&\p_r g_{00}\p_0 X^0=-\p_r g_{0r} \p_0 r,\h \p_0 X^0 \p_1 r=\p_1
X^0 \p_0 r;\ea \vspace{.5cm} \ba\nl &&\frac{\p_r g_{00}}{g_{00}}=
\frac{\p_r g_{0r}}{g_{0r}}=\frac{\p_r g_{rr}}{g_{rr}},\\ \nl
&&-\left[\p_r g_{0\alpha} (\p_0 X^0 \p_0 r -\p_1 X^0\p_1 r)+\p_r
B_{0\alpha}(\p_0 X^0 \p_1 r -\p_1 X^0\p_0 r)\right]/\p_r
g_{r\alpha}\\ \label{s03} &&=-\p_r g_{00}(\p_0 X^0 \p_0 r -\p_1
X^0\p_1 r)/\p_r g_{0r}\\ \nl &&=\p_r g_{00}\left[(\p_0
X^0)^2-(\p_1 X^0)^2\right]/\p_r g_{rr}\\ \nl &&=(\p_0 r)^2-(\p_1
r)^2 .\ea

In our second example, all string coordinates are kept fixed
except $X^0$, $X^1$ and $r$. Only to have readable final
expressions, we restrict the metric to be diagonal, and the NS-NS
field to be constant. The solutions of the corresponding equations
following from (\ref{ncsc}) and (\ref{con1}) are \ba\nl
&&\p_0 X^0=+f\p_1 r,\h \p_1 X^0=+f\p_0 r,\h \p_0 X^1=+h\p_1 r,\h
\p_1 X^1=+h\p_0 r;\\ \nl
&&\p_0 X^0=+f\p_1 r,\h \p_1
X^0=+f\p_0 r,\h \p_0 X^1=-h\p_1 r,\h \p_1 X^1=-h\p_0 r;\\
\nl
&&\p_0 X^0=-f\p_1 r,\h \p_1 X^0=-f\p_0 r,\h \p_0
X^1=+h\p_1 r,\h \p_1 X^1=+h\p_0 r;\\ \nl
&&\p_0 X^0=-f\p_1 r,\h \p_1 X^0=-f\p_0 r,\h \p_0 X^1=-h\p_1 r,\h
\p_1 X^1=-h\p_0 r,\ea where \ba\nl &&f(g,\p
g)=\left(\frac{g_{rr}\p_r g_{11}-g_{11}\p_r g_{rr}} {g_{11}\p_r
g_{00}-g_{00}\p_r g_{11}}\right)^{1/2},\\ \nl &&h(g,\p g)=
\left(\frac{g_{00}\p_r g_{rr}-g_{rr}\p_r g_{00}} {g_{11}\p_r
g_{00}-g_{00}\p_r g_{11}}\right)^{1/2}.\ea

As a consequence, one receives from here that the following
equalities are fulfilled \ba\nl &&(\p_0\pm\p_1)X^0=f(g,\p
g)(\p_0\pm\p_1)r,\h (\p_0\pm\p_1)X^1=h(g,\p g)(\p_0\pm\p_1)r.\ea
Therefore, we have obtained solutions which allow for {\it all}
string coordinates to be non-chiral.

\subsection{Solving the Boundary Conditions}
To be able to solve explicitly the boundary conditions, we assume
that $g_{MN}$ and $B'_{MN}$ are constant at $\s=0,\pi$. This is
automatically achieved if $g_{MN}$ and $B_{MN}$ depend only on
$X^a$, and the $U(1)$ field strength $F_{\mu\nu}$ is constant.
These conditions are typically realized in string theory
backgrounds, where $g_{MN}$ and $B_{MN}$ does not depend on the
coordinates along the source brane, representing the exact
solution of the {\it effective} string equations of motion, i.e.
the equations of motion in the corresponding supergravity field
theory \footnote{Here the dilaton $\Phi$ does not appear
explicitly, because we are working with the string frame metric,
in which it is included.}$ {}^{,}$ \footnote{Thus, we implicitly
imply that the D$p$-brane, on which the open string ends, is
parallel to the source (D)$p'$- brane, and $p'\ge p$.}. Moreover,
$B_{MN}$ often depends only on the radial coordinate $r$ in the
transverse to the source brane directions.

To find which non-chiral solutions of the probe string equations
of motion and constraints give also a nontrivial solution of the
boundary conditions (\ref{mbcs}) and (\ref{dbcs}), we first write
down (\ref{ncs}) in the form (compare with (\ref{vv})):
\ba\label{ncsr} &&X^M(\tau,\s)=X^M_+(\tau+\Sigma_+\s) +
X^M_-(\tau+\Sigma_-\s),
\\ \nl &&\Sigma_\pm\equiv \frac{1}{\gamma^{11}}
\left(-\gamma^{01}\pm\frac{1}{\sqrt{-\gamma}}\right)\ea Using that
the {\it background independent} solutions $X^M_\pm$ can be
expanded as \ba\nl X^M_\pm(\tau+\Sigma_\pm\s) = q^M_\pm +
\alpha^M_{0\pm}(\tau+\Sigma_\pm\s)+ i\sum_{k\ne
0}\frac{1}{k}\alpha^M_{k\pm}e^{-ik(\tau+\Sigma_\pm\s)},\ea we
represent (\ref{ncsr}) in the form \ba\label{ncsfr}
&&X^M(\tau,\s)=q^M + a^M_{0}\left(\tau -
\frac{\gamma^{01}}{\gamma^{11}}\s\right) +
\frac{b^M_{0}\s}{\gamma^{11}\sqrt{-\gamma}}\\ \nl &&+\sum_{k\ne
0}\frac{e^{-ik\left(\tau -
\frac{\gamma^{01}}{\gamma^{11}}\s\right)}}{k}
\left[ia^M_{k}\cos{\left(\frac{k\s}{\gamma^{11}\sqrt{-\gamma}}\right)}+
b^M_{k}\sin{\left(\frac{k\s}{\gamma^{11}\sqrt{-\gamma}}\right)}\right],\ea
where \ba\nl \alpha^M_{k\pm}=\frac{1}{2}\left(a^M_k \pm
b^M_k\right).\ea

Substituting (\ref{ncsfr}) into the boundary conditions, we find
the following solution of (\ref{mbcs}) and (\ref{dbcs}) \ba\nl
&&X^\mu(\tau,\s;\Sigma)=q^\mu + \left[\delta^\mu_\nu\left(\tau -
\frac{\gamma^{01}}{\gamma^{11}}\s\right) -
\left(g^{-1}B'\right)^{\mu}{}_{\nu}(q^a)\Sigma\s\right]a^\nu_{0}\\
\label{tes}&&+\sum_{k\ne 0}\frac{e^{-ik\left(\tau -
\frac{\gamma^{01}}{\gamma^{11}}\s\right)}}{k}
\left[i\delta^\mu_\nu\cos{(k\Sigma\s)} -
\left(g^{-1}B'\right)^{\mu}{}_{\nu}(q^a)
\sin{(k\Sigma\s)}\right]a^\nu_{k},\\ \nl &&X^a(\tau,\s;\Sigma)=q^a
+ \sum_{k\ne 0} \frac{e^{-ik\left(\tau -
\frac{\gamma^{01}}{\gamma^{11}}\s\right)}}{k}
b^a_k\sin{(k\Sigma\s)},\ea where \ba\nl
\left(g^{-1}B'\right)^{\mu}{}_{\nu} = g^{\mu M}B'_{M\nu} =
g^{\mu\rho}B'_{\rho\nu}+g^{\mu a}B'_{a\nu},\h \Sigma =
\frac{1}{\gamma^{11}\sqrt{-\gamma}}\in {\bf Z}_+.\ea

Thus we have showed that there exist {\it exact} solutions of the
equations of motion and constraints for the open string -
D$p$-brane system in {\it non-constant} background fields, which
are also {\it exact} solutions of the corresponding boundary
conditions. For background metric and NS-NS two-form gauge field
depending on the coordinates $X^a$, transverse to the D$p$-brane,
their explicit form is given by (\ref{tes}). This is achieved due
to the existence of the exact string solutions $X^M_\pm(w_\pm)$,
found in the previous subsection, which do not depend on the
background fields.

Let us show on an example, that there exist solutions of the
equation of motion (\ref{ncsc}) and of the constraints, which are
also solutions of the boundary conditions of the type (\ref{tes}).
To this end, consider the equations (\ref{s01}). They incorporate
two types of conditions - on the background fields and on the
string embedding coordinates. The solution for the background is
\ba\nl &&ds^2 = g(r)\left[-a^2(dx^0)^2+dr^2\right] +
2g_{0\alpha}(r)dx^0 dx^\alpha\\ \nl &&+ 2g_{r\alpha}(r)dr
dx^\alpha + g_{\alpha\beta}(r)dx^\alpha dx^\beta, \\ \label{bs}
&&B_{0\alpha}(r)=b\pm a g_{r\alpha}(r),\ea where $a$, $b$ are
arbitrary constants, and the metric coefficients are arbitrary
functions of $r$. On (\ref{bs}), (\ref{s01}) reduces to
\ba\label{ses} a\p_0 X^0(\tau,\sigma)=\pm \p_1 r(\tau,\sigma),\h
a\p_1 X^0(\tau,\sigma)=\pm \p_0 r(\tau,\sigma).\ea Replacing
(\ref{ses}) into (\ref{tes}), one obtains the following open
string solution for the background fields (\ref{bs}) \ba\nl
&&X^0(\tau,\sigma)=q^0 + i\sum_{k\ne 0}\frac{e^{-ik\tau}}{k}
\cos{(k\s)}a_k^0,\\ \nl &&X^1(\tau,\sigma)\equiv
r(\tau,\sigma)=q^r \pm a\sum_{k\ne 0}\frac{e^{-ik\tau}}{k}
\sin{(k\s)}a_k^0.\ea

Finally, in order to establish the correspondence with the known
solution of the equations of motion and of the boundary conditions
in the case of constant background fields \cite{CH98}, we make the
following restrictions:
\begin{enumerate}
\item $g_{MN}=g_{MN}(X^a)$,\h $B_{MN}=B_{MN}(X^a)$\h$\to$\\ \nl
$g_{MN}=constants$,\h $B_{MN}=constants$;

\item $g_{MN}$,\h $B_{MN}$ - arbitrary\h$\to$\\ \nl
$g_{MN}=\eta_{MN}=\mbox{diag}(-,+,\ldots,+)$,\h $B_{\mu a}=0$,\h
$B_{ab}=0$;

\item worldsheet gauge: \ba\nl\gamma^{mn}= \mbox{arbitrary
constants}\h\to \\ \nl
\gamma^{mn}=\eta^{mn}=\mbox{diag}(-,+)\h\Rightarrow\h \Sigma=1.\ea
\end{enumerate}
Under the above conditions, our solution (\ref{tes}) reduces to
the one given in \cite{CH98}.

\section{Proposal for a New Approach}
It is clear that a crucial role in treating the open string -
D-brane system in variable external fields is played by the
conditions (\ref{ncsc}), which ensure the existence of nontrivial
solutions of the type (\ref{ncs}). Actually, (\ref{ncsc}) are the
equations of motion for such type of string solutions. However,
they do not contain second derivatives. That is why, we propose to
consider them as {\it additional constraints} in the Hamiltonian
description of the considered dynamical system. So, let us compute
the resulting constraint algebra.

\subsection{Poisson Brackets}
Using the manifest expression (\ref{gm}) for the momenta, we
obtain the following set of constraints $(\p X\equiv \p X/\p\s)$
\ba\label{c0} &&I_0\equiv g^{MN}P_M
P_N-2T\left(g^{-1}B'\right)^{M}{}_{N} P_M \p X^N\\ \nl
&&+T^2\left(g-B'g^{-1}B'\right)_{MN}\p X^M\p X^N,\\ \label{c1}
&&I_1\equiv P_N\p X^N - Tg_{MK}\left(g^{-1}B'\right)^{K}{}_{N}\p
X^M\p X^N = P_N\p X^N,\\ \label{cL} &&I_L\equiv
\Gamma_{L,MN}g^{MS}g^{NK}P_S P_K
-T\left[2\Gamma_{L,MN}\left(g^{-1}B'\right)^{N}{}_{K}\right.\\ \nl
&&+\left.H_{LMK}\right] g^{MS}P_S\p X^K\\ \nl
&&+T^2\left\{\Gamma_{L,MN}\left[\left(g^{-1}B'\right)^{M}{}_{S}
\left(g^{-1}B'\right)^{N}{}_{K} - \delta^M_S
\delta^N_K\right]\right.\\ \nl &&+\left.
H_{LMS}\left(g^{-1}B'\right)^{M}{}_{K}\right\}\p X^S \p X^K .\ea
These constraints have one and the same structure. Namely, all of
them are particular cases of the expression \ba\nl &&I_J \equiv
K_J^{SK}(g,\p g)P_S P_K + S_{JK}^S(g,\p g,B',\p B')P_S\p X^K\\ \nl
&&+ R_{JSK}(g,\p g,B',\p B')\p X^S \p X^K,\ea where $J=(n,L)$, and
the coefficient functions $K_J^{SK}$, $ S_{JK}^S$ and $R_{JSK}$
depend on $X^N$ and do not depend on $P_N$. The computation of the
Poisson brackets, assuming canonical ones for the coordinates and
momenta, gives \ba\label{pba} \left\{I_{J_1}(\s_1),
I_{J_2}(\s_2)\right\}&=&
\left[M^K_{(J_1}N_{J_2)K}(\s_1)+M^K_{(J_1}N_{J_2)K}(\s_2)\right]
\p\delta(\s_1-\s_2)\\ \nl &+&C_{[J_1 J_2]}\delta(\s_1-\s_2),\ea
where $(J_1,J_2)$ and $[J_1,J_2]$ mean symmetrization and
antisymmetrization in the indices $J_1$, $J_2$ respectively.
Obviously, the algebra does not close on $I_J$. On the other hand,
the right hand side is quadratic with respect to the newly
appeared structures $M^S_J$ and $N_{JS}$. They are given by \ba\nl
M^S_J = 2K^{SN}_J P_N + S^S_{JN} \p X^N,\h N_{JS} = S^M_{JS} P_M +
2R_{JSM}\p X^M,\ea and satisfy the following Poisson brackets
among themselves \ba\nl &&\left\{M^{S_1}_{J_1}(\s_1),
M^{S_2}_{J_2}(\s_2)\right\}= \biggl[\left(K^{S_1
N}_{J_1}S^{S_2}_{J_2 N}+ K^{S_2 N}_{J_2}S^{S_1}_{J_1
N}\right)(\s_1)\\ \nl &&+\left(K^{S_1 N}_{J_1}S^{S_2}_{J_2 N}+
K^{S_2 N}_{J_2}S^{S_1}_{J_1 N}\right)(\s_2)\biggr]
\p\delta(\s_1-\s_2)+C^{S_1 S_2}_{J_1 J_2}\delta(\s_1-\s_2),\\ \nl
&&\left\{N_{J_1 S_1}(\s_1), N_{J_2 S_2}(\s_2)\right\}=
\biggl[\left(S^{N}_{J_1 S_1}R_{J_2 S_2 N}+ S^{N}_{J_2 S_2}R_{J_1
S_1 N}\right)(\s_1)\\ \nl &&+\left(S^{N}_{J_1 S_1}R_{J_2 S_2 N}+
S^{N}_{J_2 S_2}R_{J_1 S_1 N}\right)(\s_2)\biggr]
\p\delta(\s_1-\s_2)+C_{J_1 J_2 S_1 S_2}\delta(\s_1-\s_2),\\ \nl
&&\left\{M^{S_1}_{J_1}(\s_1), N_{J_2 S_2}(\s_2)\right\}=
\biggl[\left(2K^{S_1 N}_{J_1}R_{J_2 S_2 N}+
\frac{1}{2}S^{S_1}_{J_1 N}S^{N}_{J_2 S_2}\right)(\s_1)\\ \nl
&&+\left(2K^{S_1 N}_{J_1}R_{J_2 S_2 N}+ \frac{1}{2}S^{S_1}_{J_1
N}S^{N}_{J_2 S_2}\right)(\s_2)\biggr]
\p\delta(\s_1-\s_2)+C^{S_1}_{J_1 J_2 S_2}\delta(\s_1-\s_2).\ea
$M^S_J$ and $N_{JS}$ act on $P_M$ and $\p X^M$ as follows \ba\nl
&&\left\{M^{S}_{J}(\s_1), P_M(\s_2)\right\}= S^S_{JM}(\s_2)
\p\delta(\s_1-\s_2)\\ \nl &&+\left[2\p_M K^{SN}_J P_N + \left(\p_M
S^S_{JN}-\p_N S^S_{JM}\right)\p X^N\right]\delta(\s_1-\s_2),
\\ \nl
&&\left\{N_{JS}(\s_1), P_M(\s_2)\right\}= 2R_{JSM}(\s_2)
\p\delta(\s_1-\s_2)\\ \nl
&&+\left[\p_M S^N_{JS} P_N +
2\left(\p_M R_{JSN}-\p_N R_{JSM}\right)\p X^N\right]\delta(\s_1-\s_2),
\\ \nl
&&\left\{M^{S}_{J}(\s_1),\p X^M(\s_2)\right\}= 2K^{SM}_{J}(\s_2)
\p\delta(\s_1-\s_2)\\ \nl &&-2\p_N K^{SM}_J \p
X^N\delta(\s_1-\s_2),
\\ \nl
&&\left\{N_{JS}(\s_1),\p X^M(\s_2)\right\}= S^{M}_{JS}(\s_2)
\p\delta(\s_1-\s_2)-\p_N S^{M}_{JS} \p X^N\delta(\s_1-\s_2).\ea
Actually, $I_J$ can be expressed in terms of $M^S_J$ and $N_{JS}$ as
\ba\nl I_J = \frac{1}{2}\left(M^K_J P_K + N_{JK} \p X^K\right).\ea

Let us now see how from the above {\it open} algebra the
{\it closed} algebra of the constraints arises.
For the gauge generators $I_n$, we have
\ba\nl I_0 :
&&M^M_0=2\left[g^{MN}P_N-T\left(g^{-1}B'\right)^{M}{}_{N}\p X^N\right],\\ \nl
&&N_{0M}=2T\left[\left(B'g^{-1}\right)_{M}{}^{N}P_N +
T\left(g-B'g^{-1}B'\right)_{MN}\p X^N\right];\\ \nl
I_1 :
&&M^M_1=\p X^M,\h N_{1M}=P_M.\ea
Inserting these expressions in (\ref{pba}) one obtains
\ba\nl
&&\left\{I_{0}(\s_1), I_{0}(\s_2)\right\} =
(2T)^2\left[I_1(\s_1)+I_1(\s_2)\right]\p\delta(\s_1-\s_2),\\ \label{va}
&&\left\{I_{1}(\s_1), I_{1}(\s_2)\right\} =
\left[I_1(\s_1)+I_1(\s_2)\right]\p\delta(\s_1-\s_2),\\ \nl
&&\left\{I_{0}(\s_1), I_{1}(\s_2)\right\} =
\left[I_0(\s_1)+I_0(\s_2)\right]\p\delta(\s_1-\s_2).\ea
The equalities (\ref{va}) reproduce the known result stating that the string
constraint algebra in a gravitational and 2-form gauge field background
coincides with the one in flat space-time \cite{AO87}.

The Poisson bracket between $I_1$ and $I_L$ closes on $I_L$ and is given by:
\ba\nl \left\{I_{1}(\s_1), I_{L}(\s_2)\right\} =
\left[I_L(\s_1)+I_L(\s_2)\right]\p\delta(\s_1-\s_2).\ea
The remaining brackets, $\left\{I_{0}(\s_1), I_{L}(\s_2)\right\}$ and
$\left\{I_{L_1}(\s_1), I_{L_2}(\s_2)\right\}$, are of the general type
(\ref{pba}).

\subsection{Lagrangian Picture}
In order to realize from the Lagrangian approach viewpoint what
are the consequences of our idea to include another set of
constraints in the string Hamiltonian for a {\it certain type of
solutions}, now we are going to find the corresponding Lagrangian
density. To be able to compare the results step by step with the
usual case and clearly see the differences, we will first going
through the procedure namely in this case.

We start with a Hamiltonian density, which is a linear combination
of the constraints $I_0$ and $I_1$, given by (\ref{c0}) and
(\ref{c1}) respectively \ba\nl \mathcal{H}=\la^0 I_0 + \la^1
I_1.\ea Then we obtain the corresponding Hamiltonian equations of
motion for the coordinates $X^M$ \ba\nl (\p_0 - \la^1\p_1)X^M =
2\la^0g^{MN}\left(P_N -TB'_{NK}\p_1 X^K \right).\ea From here, we
express the momenta $P_M$ as functions of $\p_m X^N$ \ba\label{sM}
P_M =\frac{g_{MN}}{2\la^0}(\p_0 - \la^1\p_1)X^N +T B'_{MN}\p_1
X^N.\ea By using (\ref{sM}), the constraints $I_0$ and $I_1$ can
now be rewritten as \ba\label{c0x} &&I_0
=\frac{1}{(2\la^0)^2}g_{MN}(\p_0 - \la^1\p_1)X^M (\p_0 -
\la^1\p_1)X^N\\ \nl &&+ T^2 g_{MN}\p_1X^M \p_1X^N,\\ \label{c1x}
&&I_1 =\frac{g_{MN}}{2\la^0}(\p_0 - \la^1\p_1)X^M \p_1 X^N.\ea The
computation of the Lagrangian density gives \ba\label{ll}
\mathcal{L}&=& P_N \p_0 X^N - \la^0 I_0 - \la^1 I_1 \\ \nl
&=&\frac{1}{4\la^0}g_{MN}(\p_0 - \la^1\p_1)X^M (\p_0 -
\la^1\p_1)X^N \\ \nl &&+ T B'_{MN}\p_0 X^M\p_1X^N - \la^0T^2
g_{MN}\p_1X^M \p_1X^N.\ea

The equations of motion for $\la^m$, obtained from (\ref{ll}),
give the constraints (\ref{c0x}) and (\ref{c1x}). The
Euler-Lagrange equations for $X^M$, based on (\ref{ll}),  are
\ba\nl &&g_{LN}(\p_0 - \la^1\p_1)(\p_0 - \la^1\p_1)X^N +
\Gamma_{L,MN}(\p_0 - \la^1\p_1)X^M(\p_0 - \la^1\p_1)X^N \\
\label{emlg} &&-(2\la^0T)^2\left(g_{LN}\p_1^2 X^N +
\Gamma_{L,MN}\p_1X^M\p_1X^N\right)\\ \nl &&- 2\la^0T
H_{LMN}\p_0X^M\p_1X^N=0.\ea

Now we are interested in non-chiral solution of the above equations of
motion and constraints of the type (\ref{ncs})
\ba\label{ncsl}  X^M(\xi^m)=X^M_+(v^\la_+) + X^M_-(v^\la_-),\ea
where $X^M_\pm$ are the two background independent solutions of the
equations (\ref{emlg}), and $v^\la_\pm$ are the variables $v_\pm$ defined in
(\ref{vv}), taken in $\la$-parametrization
\ba\nl v^\la_\pm = \tau + \frac{\s}{\la^1\pm2\la^0T} .\ea
For this type of solutions, the equations of motion (\ref{emlg}) acquire
the form (\ref{ncsc}), with $w_\pm$ replaced with $v^\la_\pm$.
With the help of the explicit expressions for the momenta (\ref{sM}), one can
further transform these equations to obtain the expressions (\ref{cL})
for $I_L$, which we would like to include as additional constraints in
the Hamiltonian, describing such type string solutions.

In order to realize this idea, we now begin with the Hamiltonian
density \ba\nl \mathcal{H}=\la^0 I_0 + \la^1 I_1 + \la^L I_L.\ea
The corresponding Hamiltonian equations of motion for the
coordinates $X^M$ are \ba\nl (\p_0 - \la^1\p_1)X^M &=&
2\la^0\left(\mathcal{M}^{-1}\right)^{MN} \left(P_N -TB'_{NK}\p_1
X^K\right)\\ \nl &-&T\la^L g^{MN}H_{LNK}\p_1 X^K,\ea where we have
introduced the notation \ba\label{dg}
\left(\mathcal{M}^{-1}\right)^{MN}\equiv g^{MN} +
\frac{\la^L}{\la^0}g^{MK}\Gamma_{L,KS}g^{SN}.\ea It follows from
here that the momenta $P_M$, as functions of $\p_m X^N$, are given
by \ba\label{sMN} P_M =\frac{\mathcal{M}_{MN}}{2\la^0}(\p_0 -
\la^1\p_1)X^N +T \mathcal{A}_{MN}\p_1 X^N,\ea where \ba\label{db}
\mathcal{A}_{MN}\equiv B'_{MN} +
\frac{\la^L}{2\la^0}\left(\mathcal{M}g^{-1}\right)_{M}{}^{K}H_{LKN}.\ea
Comparing (\ref{sMN}) with the previous expressions for the
momenta (\ref{sM}), we see that they both have the same form, but
in (\ref{sMN}) new, effective background fields $\mathcal{M}_{MN}$
and $\mathcal{A}_{MN}$ have appeared. In view of (\ref{dg}) and
(\ref{db}), they have the form of a specific {\it deformation} of
the initial background.

By using (\ref{sMN}), the constraints $I_0$, $I_1$ and $I_L$ can be
rewritten as
\ba\label{c0xN} &&I_0 =\frac{1}{(2\la^0)^2}\left(\mathcal{M}g^{-1}
\mathcal{M}\right)_{MN}
(\p_0 - \la^1\p_1)X^M(\p_0 - \la^1\p_1)X^N \\ \nl
&&+\frac{T}{\la^0}\left[\mathcal{M}g^{-1}(\mathcal{A}-B')\right]_{MN}
\p_0X^M\p_1X^N \\ \nl
&&+ T^2\left\{\left[g+(\mathcal{A}-B')\right]g^{-1}
\left[g-(\mathcal{A}-B')\right]\right\}_{MN}\p_1X^M \p_1X^N,\\
\label{c1xN} &&I_1 =\frac{\mathcal{M}_{MN}}{2\la^0}(\p_0 -
\la^1\p_1)X^M \p_1 X^N, \\
\label{cLxN} &&I_L =\frac{1}{(2\la^0)^2}\left(\mathcal{M}g^{-1}
\Gamma_Lg^{-1}\mathcal{M}\right)_{MN}
(\p_0 - \la^1\p_1)X^M(\p_0 - \la^1\p_1)X^N \\ \nl
&&+\frac{T}{\la^0}\left\{\mathcal{M}g^{-1}\left[\Gamma_Lg^{-1}
(\mathcal{A}-B')
-\frac{1}{2}H_L\right]\right\}_{MN}\p_0X^M\p_1X^N \\ \nl
&&-T^2\left\{(\mathcal{A}-B')g^{-1}\left[\Gamma_Lg^{-1}(\mathcal{A}-B')
-H_L\right] + \Gamma_L\right\}_{MN}\p_1X^M \p_1X^N,\ea
where
\ba\nl (\Gamma_L)_{MN}\equiv \Gamma_{L,MN}\h\mbox{and}\h
(H_L)_{MN}\equiv H_{LMN}.\ea
The computation of the corresponding Lagrangian density gives
\ba\label{llN} \mathcal{L}^{new}&=&
P_N \p_0 X^N - \la^0 I_0 - \la^1 I_1 - \la^L I_L\\ \nl
&=&\frac{1}{4\la^0}\mathcal{M}_{MN}(\p_0 - \la^1\p_1)X^M
(\p_0 - \la^1\p_1)X^N \\ \nl
&&+T\mathcal{A}_{MN}\p_0 X^M\p_1X^N- \la^0T^2 \Delta_{MN}\p_1X^M \p_1X^N,\ea
where
\ba\nl \Delta_{MN}\equiv \left\{2g -
\left[g+(\mathcal{A}-B')\right]\mathcal{M}^{-1}
\left[g-(\mathcal{A}-B')\right] \right\}_{MN}.\ea

Comparing (\ref{llN}) with (\ref{ll}), we see that both $\mathcal{L}$
and $\mathcal{L}^{new}$ are of the same form, but in (\ref{llN}) we have
new, effective antisymmetric background field $\mathcal{A}_{MN}$, and {\it two}
new, effective background "metrics" - $\mathcal{M}_{MN}$ and $\Delta_{MN}$.

The equations of motion for the Lagrange multipliers $\la^m$,
$\la^L$, obtained from (\ref{llN}), give the constraints
(\ref{c0xN}), (\ref{c1xN}) and (\ref{cLxN}). The Euler-Lagrange
equations for $X^M$, following from (\ref{llN}), are
\ba\label{emlgN} &&\mathcal{M}_{LN}(\p_0 - \la^1\p_1)(\p_0 -
\la^1\p_1)X^N\\ \nl &&+ \Gamma^{\mathcal{M}}_{L,MN}(\p_0 -
\la^1\p_1)X^M(\p_0 - \la^1\p_1)X^N \\ \nl
&&-(2\la^0T)^2\left(\Delta_{LN}\p_1^2 X^N +
\Gamma^{\Delta}_{L,MN}\p_1X^M\p_1X^N\right)\\ \nl &&- 2\la^0T
H^{\mathcal{A}}_{LMN}\p_0X^M\p_1X^N=0,\ea where the notations used
are given by the equalities \ba\nl
&&\Gamma^{\mathcal{M}}_{L,MN}\equiv
\frac{1}{2}\left(\p_M\mathcal{M}_{NL}+\p_N\mathcal{M}_{ML}
-\p_L\mathcal{M}_{MN}\right),\\ \nl &&\Gamma^{\Delta}_{L,MN}\equiv
\frac{1}{2}\left(\p_M\Delta_{NL}+\p_N\Delta_{ML}
-\p_L\Delta_{MN}\right),\\ \nl &&H^{\mathcal{A}}_{LMN}\equiv
\p_L\mathcal{A}_{MN}+\p_M\mathcal{A}_{NL}+\p_N\mathcal{A}_{LM}.
\ea

As far as we are interested in non-chiral solutions of the above
equations of motion (see (\ref{ncsl})), we insert (\ref{ncsl}) in
(\ref{emlgN}) and obtain \ba\label{ncscN}
&&\left(\mathcal{M}_{LN}-\Delta_{LN}\right)
\left[\left(\la^1-2\la^0T\right)^2\frac{d^2X_+^N}{d(v_+^\la)^2}+
\left(\la^1+2\la^0T\right)^2\frac{d^2X_-^N}{d(v_-^\la)^2}\right]\\
\nl
&&+\left(\Gamma^{\mathcal{M}}_{L,MN}-\Gamma^{\Delta}_{L,MN}\right)
\left[\left(\la^1-2\la^0T\right)^2\frac{d X_+^M}{d v_+^\la}
\frac{d X_+^N}{d v_+^\la}\right.\\ \nl &&+\left.
\left(\la^1+2\la^0T\right)^2\frac{d X_-^M}{d v_-^\la} \frac{d
X_-^N}{d v_-^\la}\right]\\ \nl
&&-2\left(\la^1-2\la^0T\right)\left(\la^1+2\la^0T\right)
\left(\Gamma^{\mathcal{M}}_{L,MN}+\Gamma^{\Delta}_{L,MN} +
H^{\mathcal{A}}_{LMN}\right)\frac{d X_+^M}{d v_+^\la} \frac{d
X_-^N}{d v_-^\la}=0.\ea Comparing (\ref{ncsc}) with (\ref{ncscN}),
we see that the latter acquires the form of the former only when
the two "metrics", $\mathcal{M}_{MN}$ and $\Delta_{MN}$, coincide.

Since part of the Hamiltonian constraints are actually our
original Euler-Lagrange equations for $X^M$ on the specified class
of solutions, we must give new interpretation of the equations
(\ref{ncscN}), different from viewing them as equations of motion.
An important observation in this direction is that the equations
(\ref{ncscN}) contain second derivatives of the original
background fields. Therefore, they can be connected to the {\it
second variation} of the (original) action (\ref{asB}) and
consequently, to the {\it sufficient conditions} for existence of
a "local" minimum of the action on the extremal surfaces. This
problem deserves separate consideration and will be studied
elsewhere.

\section{Concluding Remarks}
In this paper we considered the case of an open string ending on
D-brane in variable external fields $g_{MN}(x)$ and $B_{MN}(x)$,
in the framework of the sigma-model approach. In Sec.2 we
formulated the problem and obtained exact solutions of the string
equations of motion, constraints and boundary conditions of a
particular type. In the constant background fields limit, our
solutions reduce to a {\it generalization} of the result already
known \cite{CH98}.

In Sec.3, we investigate some of the consequences of the idea that
the conditions for existence of non-chiral solutions of the string
equations of motion in variable gravitational and NS-NS fields,
can be reinterpreted as additional constraints. In particular, we
compute the corresponding Poisson bracket algebra, which in the
most general case does not close on the initial constraints, but
is quadratic with respect to {\it two} newly appeared structures.
The classical Virasoro algebra does not changes and appears here
as a subalgebra. In the last part of Sec.3, we analyze the changes
due to the inclusion of the new constraints from the Lagrangian
approach point of view. It turns out that this results in a
specific {\it deformation} of the initial background.


\end{document}